# Elucidation of differential response networks from toxicogenomics data.

Z. Dezso*, R. Welch*, V. Kazandaev*, A. Naito[#], J. Fuscoe[#], C. Melvin[#], Y. Dragan[#], Y. Nikolsky*, T. Nikolskaya*,[$], A. Bugrim*

*GeneGo, Inc., 500 Renaissance Drive, #106, St. Joseph, MI 49085

[#]U.S. Food and Drug Administration, National Center for Toxicological Research (NCTR) 3900 NCTR Rd., HFT-130 Jefferson, AR 72079

[$] Vavilov Institute of General Genetics, Russian Academy of Sciences, 3 Gubkina Str, Moscow, Russia

**Abstract**

We describe a novel approach to the analysis of toxicogenomics data and elucidation of biological networks affected by drug treatments. In this method approximately 15,000 linear pathway modules were generated from manually assembled pathway maps from MetaCore™ (GeneGo, Inc.). Microarray expression data from livers of rat exposed to phenobarbital, mestranol and tamoxifen were mapped onto these modules. Using different analytical techniques we have identified sets of "differential" pathways featuring highly correlated expression among multiple repeats of the same treatment while showing strong anti-correlation across different treatments. Network modules distinguishing chemical treatments were re-assembled based on these pathways. Unlike traditional statistical and clustering procedures in expression profiling, our method takes into account both network connectivity and gene expression in the course of the analysis. We demonstrate that it enables identification of important cellular mechanisms involved in drug response that would have been missed by the analysis based on individual gene expression profiles.

**Introduction**

Understanding the effects of treatment by a compound on cellular pathways and networks is increasingly important in the evaluation of a drug's safety [1,2,3,4]. Current analytical procedures in toxicogenomics are focused on statistical analysis of expression patterns, aiming at identification of small sets of genes (gene signatures) whose expression is significantly altered by treatment, and which are the most characteristic for a certain treatment [5,6]. Gene signatures are typically 50-100 genes large, and are generated using unsupervised clustering and supervised pattern matching algorithms [7,8,9]. These methods are reported to have high predictive power within the experimental set-up [10,11,12]. Based on an assumption that similar mechanisms of toxicity will evoke similar patterns of gene expression, one can deduce potential toxicity of a compound by comparing the experimentally determined signature with a library of standard signatures induced by xenobiotics with known toxicity. Some of the reference databases such as CEBS [13] and EDGE [14] are publicly available and others form the basis for commercial toxicity evaluation services by companies such as Iconix and GeneLogic. In parallel, data standards and data management systems for toxicogenomics data are being developed [15]. However, despite the obvious high utility and "easy of use" appeal, clinical application of toxicity gene signatures is limited, due to poor cross-platform, cross-experiment reproducibility [16] and functional inconsistency within signature gene content [17]. Functional analysis tools applied in drug response studies so far [2,3,18], have been limited to "enrichment analysis" in process and pathways ontologies such as GO [19] and KEGG [20]. Recently, new studies incorporate network topology in the analysis of high-throughput datasets [21,22,23, 24, 25, 26,27]. For instance, network component analysis [21], a data decomposition method based on the connectivity of the regulatory "interactome" was applied for reconstruction of transcription factor co-operation in cell cycle. Unlike "static" enrichment analysis in pre-set categories, the network-based approach is dynamic, with functional modules uniquely generated for the dataset under investigation. They also provide higher resolution at the level of direct binary interactions between proteins, genes and compounds. Previously, we have shown that interconnected protein interaction modules (signature networks) can be used as functional descriptors for compound response *in vitro* [28]. Recent case studies have also demonstrated distinct gene expression response networks for aprepitant, artemisinin and its analogs, and trovofloxacin [29].

What is most critical here is that the statistical and functional methods outlined above rely on a pre-selection of differentially expressed genes based on a fold change and/or p-value. This common wisdom selection is ubiquitously applied, because statistical thresholds are believed to enhance the validity of expression studies and current analytical methods cannot handle thousands of unfiltered data points. However, such pre-selection leaves the bulk of the expression profile beyond the scope of further functional analysis and thus may substantially change the results. In current work, we reconstruct treatment-specific networks from unfiltered expression data. Unlike the traditional workflow of pre-selecting individual genes based on expression followed by the mapping of these genes onto predefined static pathways, we have identified the interconnected modules of a global protein network with correlated response to drug action. The approach consists of the generation of a large number of relatively small, redundant sub-graphs (pathway modules), mapping of the unfiltered treatment-specific expression data on them, followed by a statistical scoring for identification of modules that are most perturbed. These modules are then combined into larger network clusters of correlated drug response, using a combination of biological and topological criteria. We have applied this method to a set of data generated for the response to several long-term drug treatments in livers of rat exposed to Phenobarbital, Mestranol and two doses of Tamoxifen [30,31].

**Results and Discussion**

Our objective was to elucidate functional differences between response to phenobarbital, tamoxifen and mestranol and to compare these findings with traditional statistical methods and with what is known about the effects of these drugs. Phenobarbital, an anticonvulsant used for treating neurological disorders such as epilepsy, is also a potent non-genotoxic carcinogen and can promote hepatocarcinogenesis in the mice model system [32,33]. To the contrary, Tamoxifen is a well-known hormonal antineoplastic agent that inhibits cell growth via an estrogen receptor dependent and independent mechanisms. Mestranol is a semi-synthetic alkylated estradiol which directly activates estrogen receptor (ESR). ESR-β may suppress breast cancer cell proliferation and tumor formation [34] and it has been reported to inhibit cyclin D1 [35] and attenuate the P53 induced apoptosis effect [36]. Although the precise roles of ESR-α and ESR-β in liver cancer are unknown, it was shown that ESR-β induces liver cancer cell apoptosis in a ligand-dependent manner [37]. Based on this information we expected to see similar biological effects between tamoxifen and mestranol, and the effect of phenobarbital to be fundamentally different.

**Identification of descriptor pathways for drug treatments.**

First, a large set of pathway modules was generated from the collection of "canonical" pathway maps manually annotated by GeneGo. These maps are freely available via iPath (www.invitrogen.com). The pathway maps display well-understood functional blocks in cell signaling and metabolism and therefore are referred to as "canonical" pathways. The pathway modules represent linear sequences of protein interactions or metabolic reactions that originate and terminate at biologically meaningful start-points (membrane receptors or their ligands) and end-points, (transcriptional factors or their immediate targets). All possible pathway modules were generated from each map by an exhaustive depth first search algorithm. The result was a total of 14,582 modules from 145 "canonical" maps containing about 1,500 genes (see "Methods" section for details). The pathways had 4.5 steps on average and had a high degree of redundancy with every gene participating on average in 10 pathways. High redundancy was needed for "fine-tuning" the selection of network modules with correlated patterns of activity. Indeed, if every gene is included in multiple pathways, most patterns of network activity can be closely approximated as a certain combination of these pathways. Pathways that were used represent cellular cascades that were described experimentally as working units, and therefore are more likely to occur in their entirety, at least in some cell types and processes, than other types of network modules.

In the next step, we prioritized the pathways in terms of relevance to treatment-specific gene expression datasets. The expression data measured as fold change (treated versus untreated samples) was mapped onto the complete set of pathways by matching gene identifiers. Relative *distances* between samples representing repeats or different treatments were then calculated in the space of gene expression of individual pathways using both Pearson correlation and Euclidian distance metrics. Euclidian distance is sensitive to changes in amplitude of gene expression, while Pearson distance should also identify pathways with small but strongly correlated differences. The overlap between the results produced with two metrics should be substantial, however we also expect to find pathways uniquely identified by each technique. We have selected pathways identified by Euclidian distance for further functional and statistical analysis.

The distances were generated for each qualified pathway among 20 expression datasets: four treatments, five repeats for each treatment. We considered a pathway "qualified" if it had at least three genes with measured expression in each of the 20 expression datasets. The pathways with fewer than three genes with expression were excluded from further analysis. The distance matrices were generated for a total of 13,151 qualified pathways. The matrices were initially analyzed using one-way ANOVA to identify the pathways for which the distances between samples representing different treatments were significantly larger than the distances among the samples representing repeats of the same treatment. In order to perform ANOVA, the distances were arranged in ten groups: four groups corresponding to the distances within the four treatments and six groups for the distances across six treatment pairs. In order to correct for multiple comparisons we applied the False Discovery Rate correction procedure [38] with an experiment significance level of 0.01.

The pathways selected by ANOVA were subjected to the t-test to identify specific treatment pairs for which selected pathways are significantly different. Based on the t-test, we selected "descriptor" pathways as being the most differentiating between pairs of treatments. The largest number of descriptor pathways were identified for the pair phenobarbital - tamoxifen (Table 1). Remarkably, there were no pathways that differentiated between tamoxifen and mestranol responses and between the two concentrations of tamoxifen. These results are consistent with biological differences and similarities of these drugs' effects mentioned above. Indeed, phenobarbital is known to induce cell proliferation and tumor growth while tamoxifen is an estrogen antagonist with clear anti-growth effect. On the other hand, mestranol is a synthetic analog of estradiol and similar to tamoxifen in its ability to interact with estrogen receptor. Application of Pearson distance yielded more differentiating pathways in each comparison than the Euclidian distance. This indicates that differences between treatments may manifest themselves not only in amplitude of expression change along a pathway but also as small coordinated changes in expression for a whole pathway. Such pathways are missed by the analysis based on Euclidian distances but are identified in Pearson correlation distance matrix.

In the next step, we exported all genes from the descriptor pathways and organized them as non-redundant gene lists characteristic for each pair of treatments. The number of genes differentiating between treatments are shown in Table 1. The relatively small number of genes compared to the number of pathways is an indication that the descriptor pathways share a large number of common genes. We compared the intersection between the gene lists differentiating between treatments (Fig. 1). 51% of the genes from phenobarbital-mestranol list were common with the phenobarbital-tamoxifen (low dose) pair and 79% genes with the phenobarbital-tamoxifen (high dose) list (Fig. 1). The descriptor genes for phenobarbital and the two concentrations of tamoxifen pairs highly overlapped (close to 90%), the descriptor genes for phenobarbital tamoxifen (low dose) representing a subset of the descriptor group for the Phenobarbital-Tamoxifen (high dose) pair. Our observations are consistent with the degree of similarity between treatments and make biological sense. Indeed, the most similar treatments, those for the two doses of the same drug (tamoxifen) show no significant differences in affected pathways. Furthermore, there were no descriptor pathways for mestranol-tamoxifen pair. Indeed, mestranol and tamoxifen, have similar biological targets (both are ligands for the estrogen receptor), while phenobarbital is distinct from this pair, both structurally and by mode of action.

**Differential response networks.**

The characteristic gene lists for treatment pairs were used as input files for the generation of biological networks in MetaCore [28]. Two gene lists were generated for each treatment pair via ANOVA test of the sample-sample matrix. The networks were built by direct interactions (DI) algorithm which allows visualization of only the edges which correspond to direct physical interactions between the input nodes; in this case the protein products of the genes from the characteristic lists (Fig. 2). The interconnected parts of the networks contained 99 and 37 nodes for Phenobarbital-tamoxifen and for Phenobarbital-mestranol pairs, respectively. To test the significance of the results we randomly changed the labels of the genes across the entire expression data set and repeated the analysis 100 times. As we expected the random mixing of labels broke the association between expression profile and network topology, leading to a smaller number of differentiating pathways (genes). We did not find *any* gene differentiating between most of the treatment pairs and identified a much lower number of genes differentiating between Phenobarbital and the two different concentrations of tamoxifen (both corresponding to $p=0.05$) We also investigated the topological properties of the networks. Some of the properties of differential networks showed significant differences in comparison to networks generated from a set of randomly picked pathways (Table 2).

The resulting networks were overlaid with expression data presented as a ratio between treated and untreated liver expression (Fig. 2). Expression of the majority of the genes on these networks are anti-correlated between treatments. The fold change difference for the genes on the networks varied between 1% and 65% and standard t-test p-values of individual data points between 10-e4 and 0.97. Therefore, a large fraction of these genes (55% for p<0.1 threshold) would have been excluded from the analysis by any conventional microarray analysis procedure based on statistical significance of individual genes. We have also identified the network modules specific for certain functional processes (Figure 2c), by selecting subsets annotated to particular biological processes as defined by Gene Ontology [19].

The networks can also be characterized and compared based on their global topology features such as the number of nodes, presence of highly connected nodes (hubs), in- and out-going edges, transcription factors, and receptors involved in the networks etc. The top 10 hubs defined by the number of connections and the top 10 transcription factors are shown in Table 2.

We also calculated average degree, average clustering coefficient, average shortest path and centrality of the nodes for three types of networks: the global interconnected cluster based on all interactions in the MetaCore™ database, the network of protein content from MetaCore™ canonical maps connected through direct interactions; and differential response networks. We calculated the network properties in the global protein interaction network and averaged the quantities over the proteins identified as differentiating between the treatments. The network for proteins gleaned from our maps feature higher than average degree (in- and out degree), clustering coefficient, centrality and shorter than average shortest path than the global network. This was expected, as maps are generated from the pathways experimentally shown to be the main signaling conduits and metabolic fluxes in human cells. However, the substantial differences in some of the topological properties between drug response networks and networks built from proteins on our maps were surprising (Table 2). In order to assign statistical significance to these differences we calculated p-values of networks created from a set of randomly chosen pathways (same as the real number of differentiating pathways). The drug response network for Phenobarbital and tamoxifen (low dose) featured a disproportionately high fraction of "hubs", manifested in significantly higher average node degree for both incoming and outgoing interactions. Similar tendencies were reported for protein interaction networks in yeast for essential proteins and toxicity modulation proteins [39,40]. Interestingly, the descriptor network for phenobarbital vs. tamoxifen (high dose) pair features a slightly lower clustering coefficient, probably reflecting a different drug response mechanism. The shorter than average shortest path may be associated with a network optimized for a quick response time.

**Functional analysis of signature networks.**

Using MetaCore™ tools, we performed functional enrichment analysis for the gene content of differential response networks' using canonical pathway maps and Gene Ontology (GO) processes. The distribution of the top 10 canonical maps and GO processes for two signature networks (for mestranol vs. phenobarbital and tamoxifen vs. phenobarbital) is shown in Table 3. The statistical significance of functional maps was evaluated by Monte-Carlo procedure as described in "Methods" section. In this procedure we have simulated random selection of sets of pathway modules with subsequent mapping onto maps. We choose this procedure over hypergeometric distribution to account for the fact that genes in differential response networks are always selected as part of a pathway module, rather than individually. The networks for phenobarbital-tamoxifen and phenobarbitral-mestranol pairs displayed quite different enrichment profiles despite the fact that they share one of the counterpart treatments (phenobarbital). We suggest that the two networks reflect different functional aspects of drug action mechanism. Most of the top scoring maps distinguishing phenobarbital and tamoxifen are related to cell cycle regulation (Table 3). Remarkably, all of the top four GO categories for this treatment pair are also related to cell cycle or its elements. This indicates that our method is robust in identifying cell proliferation effect of phenobarbital when compared to tamoxifen. On the other hand, differences between phenobarbital and mestranol are most prominent in how these drugs affect signal transduction such as MAP-kinase pathway, growth-factor signaling, etc. We believe that this is mostly due to the prominent hormonal role of mestranol. This is further confirmed by top GO processes related to phenobarbital-mestranol network, which include signal transduction, phosphorylation and cell communication.

Closer examination of top scoring maps reveals some interesting mechanistic observations. For instance, they show opposite effects of phenobarbital and tamoxifen on major elements of the cell cycle. These effects are very modest in magnitude, rarely exceeding 40% change in gene expression but very consistent for most of the pathway elements. Figure 3a shows an overview map of mitosis initiation with gene expression data for all three drugs mapped on it. The map is one of the top-scored maps for phenobarbital-tamoxifen comparison and has four transcription factors serving as pathway inputs: c-Myc, AP-2A, USF1 and FOXM1. All four are up-regulated by phenobarbital treatment and down-regulated by tamoxifen and mestranol treatments. The major target of these transcription factors is cyclin B1, which in turn activates CDK1 – a key regulator of entry into mitosis. Earlier, it was shown that treatment of c-Myc transgenic mice with Phenobarbital indeed results in tremendous acceleration of neoplastic development in the liver compared with non-treatment of c-Myc mice or treatment of wild-type mice with Phenobarbital [33]. The role of FOXM1 in liver carcinomas is also documented [41].

Interestingly, this map shows fairly similar expression profiles for tamoxifen and mestranol, yet only two genes from phenobarbital-mestranol differential response network are present on it. This serves as an early indication that our technique is capable to detect pathways with small but consistent quantitative variations in gene expression which may result in phenotype divergence. Indeed despite that all gene expression variations appear equally small strong association of phenobarbital-tamoxifen network with cell cycle points to major differences these drugs might have had on liver cell proliferation if administered for longer time or at higher doses. Majority of descriptive genes identified by our method would have been rejected by any "rule of thumb" constraints used in microarray analysis. For example, a mild constraint of fold change > 1.4 and $p < 0.1$ leaves no "differentially expressed" genes on the mitosis map  Moreover, if statistical constraints are imposed on the gene expression data prior to mapping, all maps presented above are scored low, and therefore

11are not even considered as a relevant functional descriptor between the datasets. We applied conventional t-test (Phenobarbital vs. Tamoxifen (low dose)) to identify the genes differentiating between treatment pairs. We note that after we applied the False Discovery Rate approach at significance level 0.01 we found only 9 genes differentiating between the two treatment pairs. Further, a less stringent cutoff of $p<0.01$ and no multiple testing correction yielded 330 differentiating genes. The functional analysis of these genes showed that there were only *two* significantly enriched pathway maps with no clear connection with the treatment pairs (Ral1A regulation pathways and A2BR signaling via G-beta/gamma dimmer). The top maps identified by our method were not significantly enriched in the differentiating genes based on the t-test, the p-values being larger than any reasonable threshold ($p>0.1$ for most top maps). For example, the "start of mitosis" map has only a single protein, Cyclin B1 (Fig 3a, circled in green) which is differentiating between phenobarbital and tamoxifen based on the t-test.

We compared our method with the traditional hierarchical clustering approach that relies on the fold changes of individual genes. We calculated the Pearson distance between the log ratios of gene expression profiles of individual genes and performed standard hierarchical clustering [42] on all genes with expression data for each of 20 datasets which are present on the generated pathways. The genes not present on pathways were excluded. Clustering failed to clearly identify gene sets with any difference on a pathway level, which is well consistent with literature. To illustrate this point further, we marked genes identified by our pathway-based method with bars of different color based on treatment comparisons (Fig. 3b). None of these bars seem to form any cohesive groups on the clustering map, suggesting that clustering cannot identify descriptor genes which form cohesive differential-response networks.

**Methods**

**Gene expression datasets.**

Gene expression data were produced as part of the program to study effects of long-term drug exposure at the National Center for Toxicological Research [33]. Female rats were administered the AIN-76A basal diet or a diet containing the drugs mixed into the basal diet at 500 mg Phenobarbital per kg diet, 2 mg mestranol per kg diet, and tamoxifen at either 250 mg per kg diet (designated as low) or 500 mg per kg diet (designated as high). Overall, five treatment groups were used in the experimental design: control, phenobarbital, mestranol, and 2 doses of tamoxifen. RNA from rat livers was isolated and used for the microarray experiment. All samples were co-hybridized with Stratagene universal rat reference RNA. The design is a 2-color common reference design experiment with the samples labeled with Cy5 and the reference always labeled with Cy3. The 10,000 rat oligonucleotide probes were purchased from MWG and printed onto poly-L-lysine slides at NCTR Center for Functional Genomics. For each treatment group data for five biological repeats were assayed. For further analysis, we have computed log-ratio for treated vs. untreated animals according to the following formula:

$\log_2$ (fold change) = $\log_2(T/C)$ = $\log_2(T/R)$ - $\log_2(C/R')$, where

T = signal from treated sample

C = signal from control (untreated sample)

R = signal from reference sample for treatment

R' = signal from reference sample for control

**Generation of the sample-sample distance matrix.**

After mapping gene expression data onto pre-computed pathway modules the matrix of sample-sample distances was calculated. Distances were calculated in the gene expression space of every individual pathway module.

We used both Euclidian distance and Pearson correlation (in Supplementary Materials) for calculating two distance matrices. The Pearson distance between samples $x$ and $y$ for a pathway of n genes is calculated as $d_{x,y} = 1 - \frac{1}{n}\sum_{i=1}^{n}\left(\frac{x_i - \bar{x}}{\sigma_x}\right)\left(\frac{y_i - \bar{y}}{\sigma_y}\right)$, where $x_i$ and $y_i$ are the gene expression log ratios for genes $i$ in sample $x$ and $y$, and $\bar{x}, \bar{y}, \sigma_x, \sigma_y$ are the mean values and respective standard deviations. Similarly, the Euclidian distance is calculated as $d_{x,y} = \sqrt{\sum_{i=1}^{n}(x_i - y_i)^2}$. After calculating all distances, the procedure results in two matrices of sample-sample distances. Each row of these matrices corresponds to a pathway module and each column represents a sample pair. Thus for 20 samples (5 repeats for each treatment) the size of each matrix is 13,151X190.

**Identifying condition-specific pathways modules based on ANOVA and t-test.**

Calculated sample-sample distances were separated into 10 groups. Four groups correspond to individual drug treatments and contain distances between repeats of the same treatment. Six groups contain distances from individual repeats of one treatment to individual repeats of another for all six possible treatment pairs. Using these groups we performed one-way ANOVA analysis to reveal pathways for which there are groups with significantly different distance distributions. Alternatively, as the normality requirement for ANOVA may not necessarily hold for sample-sample distance distribution, we applied the Kruskal-Wallis method which gave identical results. After ANOVA pathways were rank ordered according to calculated p-values and the False discovery rate correction was applied, only the pathways for which $p_k < α*k/m$ were selected for further analysis. Here $m = 13,151$ number of pathways tested, $k$ is the rank of individual pathway, and $α = 0.01$ is experiment significance level. For the pathways selected by ANOVA and subsequent FDR correction, the pair-wise t-test was performed to determine which specific groups of distances are significantly different. For the t-test, a 0.01 significance-level was used and the Bonferroni correction was applied. As the number of comparisons for each pathway is twelve (two comparisons for each of the six treatment pairs) the corrected significance level we used is 8.3e-4. The selected pathways were grouped into "clusters" based on the t-test results and we selected the ones which clearly distinguish between two treatments. Specifically, we selected those for which there existed particular pair of treatments so that distances among repeat samples for the same treatment are significantly smaller than the inter-treatment distances. For instance, the pathways that distinguish mestranol from tamoxifen treatment had to satisfy the condition that both distances between mestranol repeats and distances between phenobarbital repeats are statistically smaller than the mestranol-phenobarbital distances.

**Evaluating significance of functional enrichment.**

We evaluated statistical significance of the enrichment of functional maps by using a randomization procedure. To account for the fact that genes in our mappings are derived from "differentiating" pathway modules, our procedure started with randomly picking pathways modules rather than individual genes. Pathways were selected until we reached the same size of gene content as the size of our differential response networks. In order to achieve exactly the same number of genes, once the gene content of randomly picked pathway modules became larger than our differentiating gene set we simply corrected by retaining an appropriate subset of interconnected genes from one of the randomly picked pathway modules. We repeated this procedure 1,000,000 times, each time calculating the overlaps of a resulting set of genes with all functional maps. Next we calculated the overlaps of gene sets derived from differential response networks with every map and estimated p-values as relative frequency of instances where the random sets had higher overlap with a map than the differentiating set of genes.

We applied FDR at 0.01 significance level and identified 8 map significantly enriched for Phenobarbital and Mestranol treatments and 16 maps significantly enriched for Phenobarbital and Tamoxifen treatments.

**Topological measures**

*Degree of nodes.* The number of links connected to a node gives the node's degree. Since many real networks are directed, nodes are characterized by in and out-degree, giving the number of outgoing and incoming interactions. While calculating average degree of nodes in a network, we average over the degree of the nodes which are part of that network, but considering all interactions they have in the global network. Similarly, while calculating clustering coefficient we consider all interactions of a node in the global network.

*Average shortest path.* The shortest distance between two nodes is the number of links along the shortest path. The average shortest path is the average over the shortest paths for all node pairs in the network. When we calculate the shortest paths for a subset of nodes in the network we consider also paths crossing through nodes which are not part of the subset.

*Average clustering coefficient.* The clustering coefficient is a measure that captures to what degree node's neighbors are connected. It is defined as: $C_i = \frac{2n_i}{k_i(k_i-1)}$, where $n_i$ is the number of links among the $k_i$ neighbors of node *i*. As $k_i(k_i-1)/2$ is the maximum number of such links, the clustering coefficient is a number between 0 and 1. The average clustering coefficient is obtained by averaging over the clustering coefficient of individual nodes. A network with high clustering coefficient is characterized by highly connected sub-graphs.

*Centrality of nodes.* Centrality of a node is the number of shortest paths going through that node when we consider the shortest path between all node pairs. When there is more than one shortest path between two nodes the centrality is divided by the number of shortest paths between them. The centrality of a node (*k*) can be calculated as:

$$C_k = \sum_{i \neq j} \frac{\sigma_{ij}(k)}{\sigma_{ij}},$$

where $\sigma_{ij}$ is the number of shortest paths between *i* and *j*, $\sigma_{ij}(k)$ is the number of shortest paths between *i* and *j* which passes through *k*.

**Acknowledgements**

This work was supported by the NIH grant 1R43ES01380001 and by the DoD grant FA865005M6596.

Tables

Table 1. The number of pathways and genes differentiating between the treatment pairs. The microarray data were produced as follows: female rats were administered the AIN-76A basal diet or that diet containing the drugs admixed into the basal diet at 500 mg Phenobarbital per kg diet, 2 mg mestranol per kg diet, and tamoxifen at either 250 mg per kg diet (designated as low) or 500 mg per kg diet (designated as high).

| Treatments | Number of differentiating pathways | Number of differentiating genes |
|---|---|---|
| Phenobarbital vs. mestranol | 12 | 58 |
| Phenobarbital vs. tamoxifen (low) | 123 | 146 |
| Phenobarbital vs. tamoxifen (high) | 261 | 334 |
| Mestranol vs. Tamoxifen (low) | 0 | 0 |
| Mestranol vs. Tamoxifen (high) | 0 | 0 |
| Tamoxifen (low) vs. Tamoxifen (high) | 0 | 0 |

Table 2. Network characteristics.

A) Topological characteristics of differential drug response networks compared to global network. Values with statistically significant differences are highlighted in bold and values are calculated in the global protein interaction network and averaged over the set of differential genes. Only p-values below 0.1 are indicated in the table. B) The top 10 hubs and transcriptional factors characterized by highest connectivity calculated in the "differential" response networks (Fig. 1)

(A)

| Topological properties | Global network | Proteins from maps | Phenobarbital –Mestranol | Phenobarbital-Tamoxifen(low) | Phenobarbital-Tamoxifen (high) |
|---|---|---|---|---|---|
| Degree | 10.7 | 27.68 | 49.57 | **45.23 (p=0.004)** | 38.14 (p=0.08) |
| In-degree | 5.27 | 13.09 | 27.08 | **23.36 (p=0.004)** | 19.26 |
| Out-degree | 5.37 | 15.52 | 22.49 | **21.87 (p=0.02)** | **18.88 (p=0.05)** |
| Clustering coefficient | 0.14 | 0.14 | 0.15 | 0.14 | 0.12 |
| Shortest path | 4.42 | 3.27 | **2.33 (p<0.001)** | **2.74 (p<0.001)** | 3.04 (p=0.06) |
| Centrality ($10^4$) | 9.9 | 23.5 | 39.5 | 35.5 | 30.92 (p=0.1) |

(B)

| Phenobarbital/Tamoxifen network | | | |
|---|---|---|---|
| Hubs | Connectivity | Transcription factors | Connectivity |
| p53 | 34 | P53 | 34 |

| Hubs | Connectivity | TFs | Connectivity |
|---|---|---|---|
| CDK1(p34) | 30 | STAT3 | 18 |
| c-Src | 27 | c-Cbl | 17 |
| Shc | 24 | c-Jun | 17 |
| SHP-2 | 23 | CREB1 | 17 |
| GRB2) | 22 | E2F1 | 13 |
| JAK2 | 22 | AP-2A | 9 |
| PI3K | 19 | HIF1A | 9 |
| Phenobarbital/Mestranol network | | | |
| Hubs | Connectivity | TFs | Connectivity |
| RelA (p65) | 20 | RelA (p65) | 20 |
| NFKBIA | 19 | c-Fos | 14 |
| c-Src | 18 | c-Jun | 13 |
| c-Fos | 14 | NF-KB1 | 13 |
| EGFR | 14 | NF-KB1 (p105) | 13 |
| c-Jun | 13 | c-Rel | 10 |
| IRS-1 | 13 | NF-kB p50/p65 | 10 |
| NF-kB1 | 13 | NF-kB2 (p100) | 10 |
| Erk (MAPK1/3) | 12 | RelB | 10 |

Table 3. Functional characteristics of phenobarbital/mestranol and phenobarbital/tamoxifen differential response networks

| Pathway maps | p-value | Gene Ontology "Biological process" | p-value |
|---|---|---|---|
| Phenobarbital – tamoxifen (low) comparison | | | |
| Peroxisomal branched chain fatty acid oxidation | 4.40E-05 | cell cycle | 1.67E-08 |
| Role APC in cell cycle regulation | 8.10E-05 | regulation of cell cycle | 2.20E-08 |
| H-RAS regulation pathway | 0.000201 | regulation of progression through cell cycle | 5.91E-08 |
| The metaphase checkpoint | 0.000685 | transmembrane receptor protein tyrosine kinase signaling pathway | 3.93E-07 |
| Start of the mitosis | 0.000742 | mitotic cell cycle | 7.21E-07 |
| Propionate metabolism p.2 | 0.001004 | Mitosis | 1.22E-06 |
| Spindle assembly and chromosome separation | 0.001324 | enzyme linked receptor protein signaling pathway | 2.30E-06 |
| FGF-family signaling | 0.001448 | organ morphogenesis | 3.39E-06 |
| Saturated fatty acid biosynthesis | 0.001904 | Biopolymer modification | 6.19E-06 |
| Mitochondrial unsaturated fatty acid beta-oxidation | 0.002405 | Morphogenesis | 6.41E-06 |
| Phenobarbital – mestranol comparison | | | |
| GDNF family signaling | <1.00E-06 | signal transduction | 2.43E-11 |
| EGF signaling pathway | 3.00E-05 | cell communication | 1.35E-10 |
| MAPK cascade. Part II, Map 1 - MAP4K2-4 - ERK5 & JNK pathway | 3.40E-05 | phosphorylation | 2.19E-09 |
| H-RAS regulation pathway | 0.00013 | intracellular signaling cascade | 7.79E-09 |
| PDGF signaling via STATs and NF-kB | 0.000203 | phosphate metabolism | 1.47E-08 |
| Pleckstrin homology proteins interactions. Part III | 0.000303 | phosphorus metabolism | 1.47E-08 |
| Angiopoietin - Tie2 signaling | 0.000323 | protein kinase cascade | 2.45E-08 |
| Role of PDGFs in cell migration | 0.000365 | protein amino acid phosphorylation | 2.53E-08 |
| A2A receptor signaling | 0.000765 | biopolymer modification | 8.78E-07 |
| MAPK cascade. Part I. Map 1. ERK-related pathways | 0.001292 | anatomical structure development | 2.31E-06 |

**Figures**

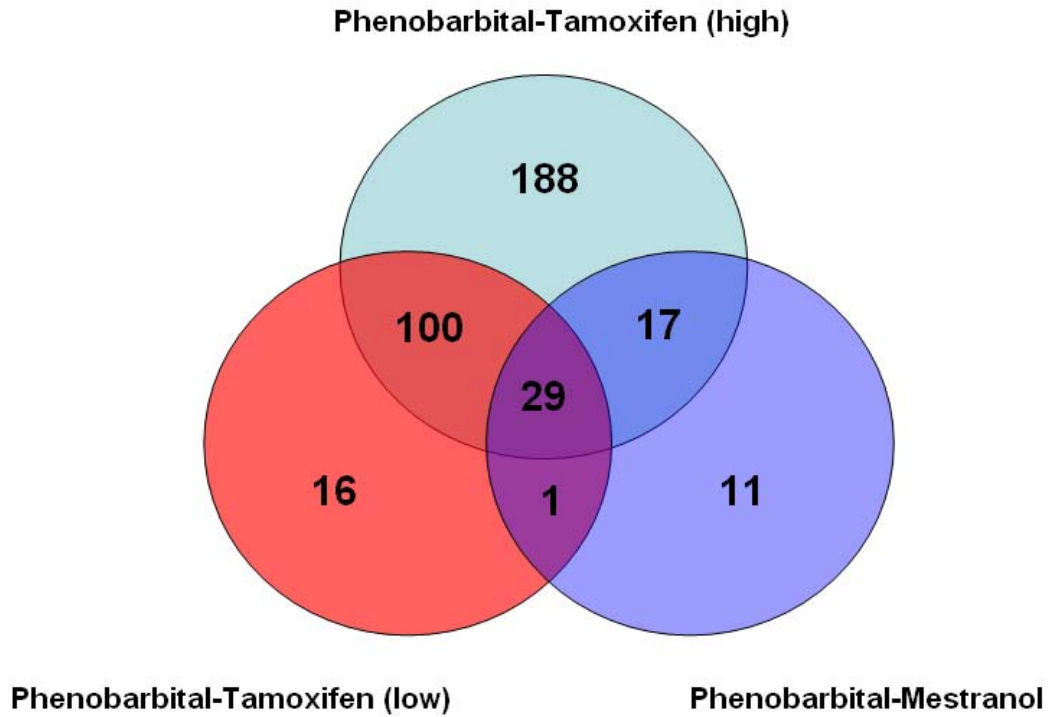

Figure 1. Characterizing the overlap between groups of genes identified as differentiating between treatment conditions. Approximately 51% of the genes from phenobarbital-mestranol list were common with the phenobarbital-tamoxifen (low dose) pair and 79% genes with the phenobarbital-tamoxifen (high dose) list. The descriptor genes for phenobarbital and the two concentrations of tamoxifen pairs highly overlap (close to 90%).

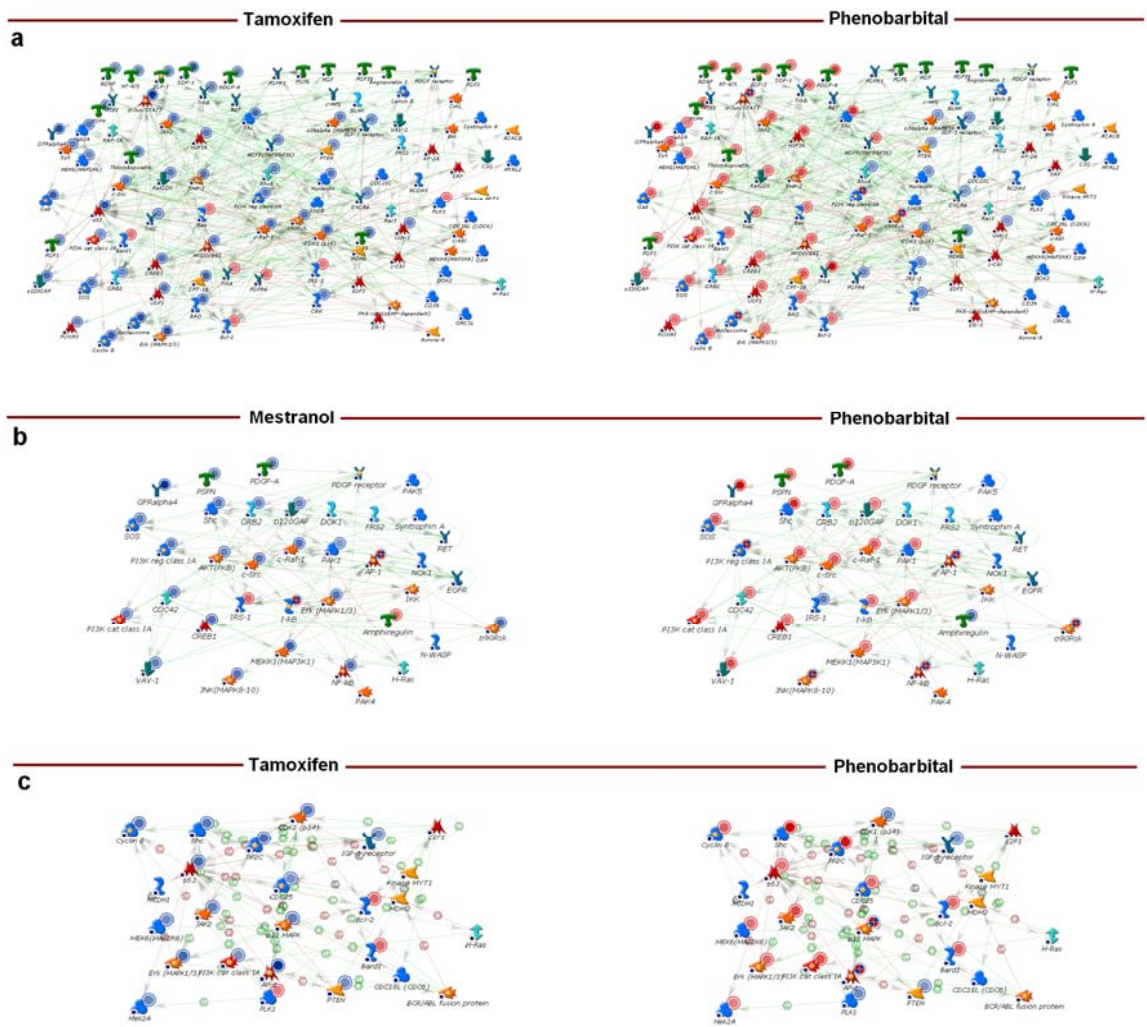

Figure 2. Differential drug-response networks are based on pathways that differentially respond to treatments. (a) Direct interaction network assembled from a group of genes from pathways differentiating mestranol and phenobarbital treatments. Relative changes in expression between treated and untreated rats are mapped (log-ratios, averaged over 5 repeats). Atop each network, we noted the treatment to which gene expression profile corresponds. (b) The network assembled from the genes extracted from pathways that distinguish phenobarbital from tamoxifen. Note that both comparisons contain significant numbers of genes that are anti-correlated between treatments. (c). A sub-graph network connecting only the genes annotated to "cell cycle" in Gene Ontology

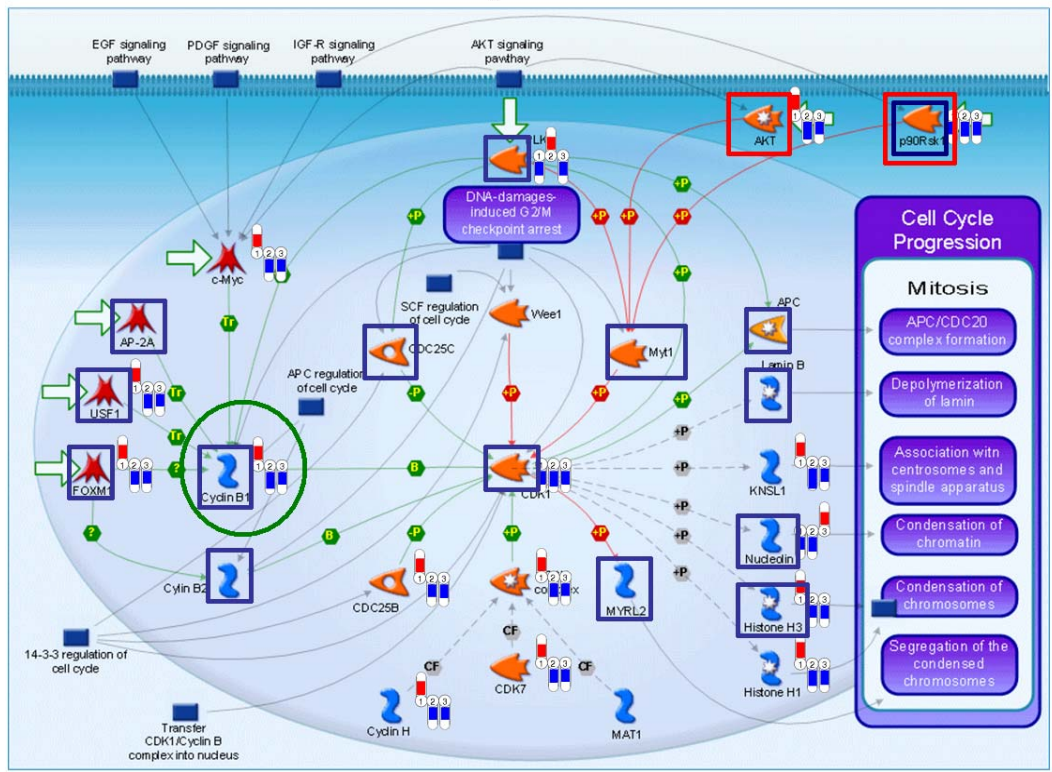

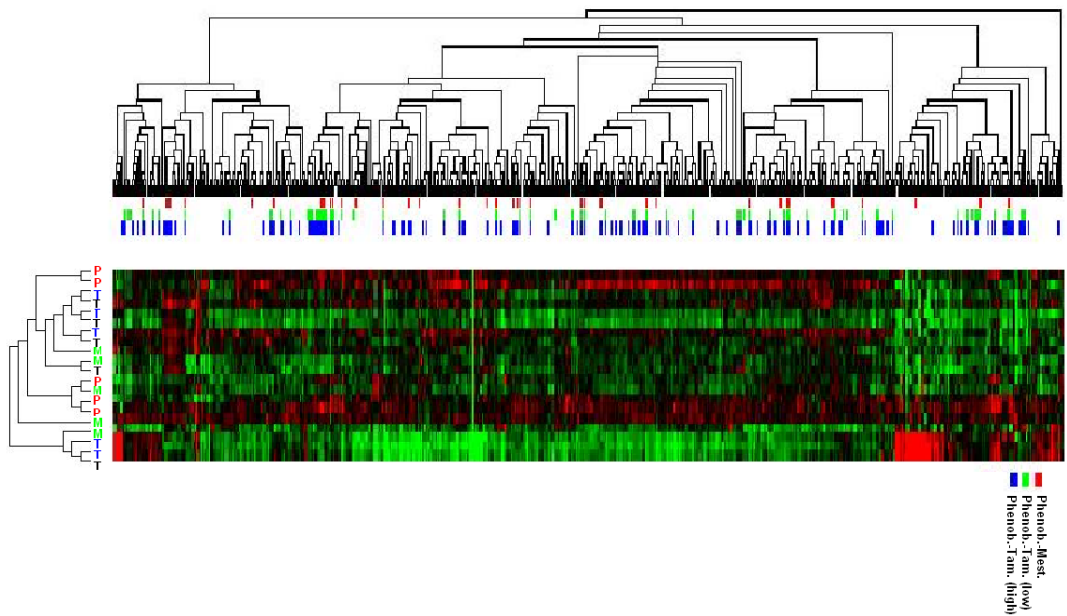

Figure 3. Differential drug response data on maps and bi-clustering of genes. (a) "Start of mitosis" map is one of top scored map for the enrichment in the set of genes from the network differentiating between tamoxifen and phenobarbital treatments. Thermometer-like icons correspond to over- (red)

and under- (blue) expression and numbered to indicate treatment to which they correspond. Expression data for phenobarbital (#1), tamoxifen (lower concentration, #2) and mestranol (#3) are mapped without pre-filtering. Note that most genes on this map are consistently under expressed for tamoxifen and over-expressed for phenobarbital treatments, resulting in a map clearly differentiating between the two responses. The green circle shows the only gene that was identified as differentiating between the two treatments by a conventional t-test. Boxes indicate proteins that are members of differential response networks for phenobarbital-tamoxifen (blue) and phenobarbital-mestranol (red). Since differential response networks are built from whole pathway modules, some of their member proteins may have no expression data associated with them.

(b) Bi-clustering of genes and treatments based on expression. Only the genes that could be mapped onto our set of pathway modules are used in clustering. Colored bars identify genes selected by our pathway-based method as differentiating between treatments: red - phenobarbital vs. mestranol, green - phenobarbital vs. tamoxifen (low), blue - phenobarbital vs. tamoxifen (high).